\documentclass[runningheads]{svmult}

\usepackage{epsf}
\usepackage{makeidx}   % allows index generation
\usepackage{graphicx}  % standard LaTeX graphics tool
\usepackage{subeqnar}  % subnumbers individual equations
\usepackage{multicol}  % used for the two-column index
\usepackage{physprbb}  % centered layout of diverse elements, etc.
\def\fcite#1{\cite{#1}}
\newcommand {\be}{\begin{eqnarray}}
\newcommand {\ee}{\end{eqnarray}}

\def\gsim{\lower0.5ex\hbox{$\stackrel{>}{\sim}$}}
\def\lsim{\lower0.5ex\hbox{$\stackrel{<}{\sim}$}}

\def\overbar{\overline}

\begin{document}
\title*{SIMP (Strongly Interacting Massive Particle) 
Search\protect\footnote{Presented by Vigdor L. Teplitz. 
To be published in the proceedings of 4th International 
Symposium on Sources and Detection of Dark Matter in the Universe (DM
2000), Marina del Rey, California, 23-25 Feb 2000. 
}}
\toctitle{SIMP (Strongly Interacting Massive Particle) Search}
\titlerunning{SIMP Search}
% allows abbreviation of title, if the full title is too long
% to fit in the running head
%
\author{Vigdor L. Teplitz\inst{1}\protect\footnote[2]{%
Address until June, 2001: 
Office of Science and Technology Policy, 
Executive Office of the President,
Washington, DC 20502.}
\and Rabindra N. Mohapatra\inst{2} 
\and Fred Olness\inst{1}
\and Ryszard Stroynowski\inst{1}
}
\authorrunning{Vigdor L. Teplitz et al.}
% if there are more than two authors,
% please abbreviate author list for running head
%
%
\institute{Department of Physics, 
Southern Methodist University,  Dallas, TX 75275
\and
Department of Physics, University of Maryland, 
College Park, MD 20742.}

\maketitle              % typesets the title of the contribution

%%%%%%%%%%%%%%%%%%%%%%%%%%%%%%%%%%%%%%%%%%%%%%%%%%%%%%%%%%%%%%%%%%%%%%%%
\begin{abstract}
We consider laboratory experiments that can detect stable, neutral
strongly interacting massive particles (SIMPs). We explore the SIMP
annihilation cross section from its minimum value (restricted by
cosmological bounds) to the barn range, and vary the mass values from
a GeV to a TeV.  We also consider the prospects and problems of
detecting such particles at the Tevatron.
\end{abstract}

%%%%%%%%%%%%%%%%%%%%%%%%%%%%%%%%%%%%%%%%%%%%%%%%%%%%%%%%%%%%%%%%%%%%%%%%
%%%%%%%%%%%%%%%%%%%%%%%%%%%%%%%%%%%%%%%%%%%%%%%%%%%%%%%%%%%%%%%%%%%%%%%%
%%%  FIGURES
%%%%%%%%%%%%%%%%%%%%%%%%%%%%%%%%%%%%%%%%%%%%%%%%%%%%%%%%%%%%%%%%%%%%%%%%
%%%%%%%%%%%%%%%%%%%%%%%%%%%%%%%%%%%%%%%%%%%%%%%%%%%%%%%%%%%%%%%%%%%%%%%%
%%%%%%%%%%%%%%%%%%%%%%%%%%%%%%%%%%%%%%%%%%%%%%%%%%
\def\figthree
{
\begin{figure}[t]
 \begin{center}
 \leavevmode
 \epsfxsize=0.50\hsize \epsfbox{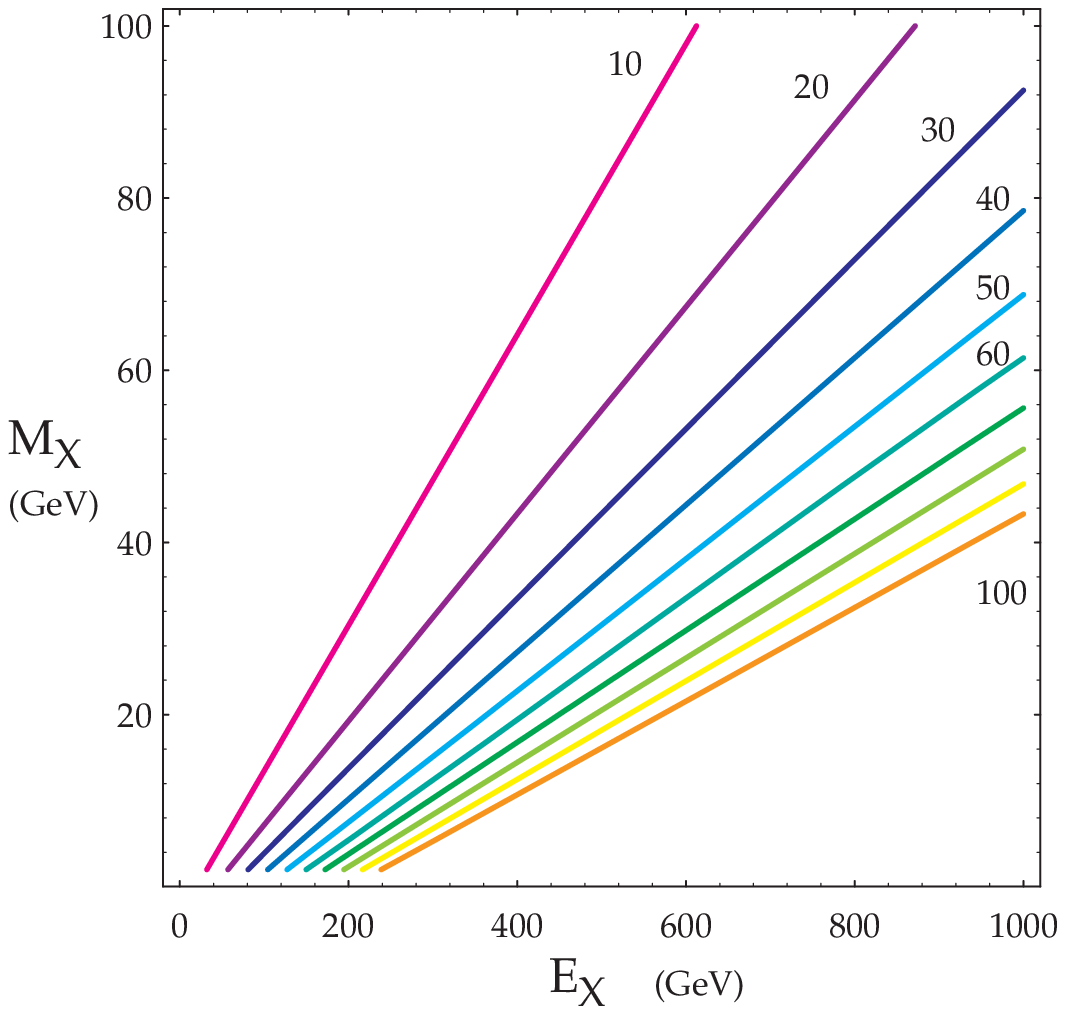}
 \end{center}
% \tightenlines
 \caption{ 
Contours for energy loss as a function of $\{M_X,E_X \}$. 
The contours displayed are in steps of 10 GeV. 
 }
 \label{fig:three}
\end{figure}
}
%%%%%%%%%%%%%%%%%%%%%%%%%%%%%%%%%%%%%%%%%%%%%%%%%%%%%%%%%%%%%%%%%%%%%%%%
%%%%%%%%%%%%%%%%%%%%%%%%%%%%%%%%%%%%%%%%%%%%%%%%%%
\def\figtab
{
\begin{figure}[t]
 \begin{center}
 \leavevmode
 \epsfxsize=0.95\hsize \epsfbox{figtab.eps}
 \end{center}
% \tightenlines
 \caption*{ 
$M_X$ (vertical) is in units of GeV, and $\sigma_{XN}$ (horizontal) is in units of $mb$.  
 Table entries are $\log_{10}(1/f)$, and
 the $-$ indicates those cases for which $X$ does not bind at all.
 }
 \label{fig:tab}
\end{figure}
}
%%%%%%%%%%%%%%%%%%%%%%%%%%%%%%%%%%%%%%%%%%%%%%%%%%%%%%%%%%%%%%%%%%%%%%%%
%%%%%%%%%%%%% END DEFINITIONS %%%%%%%%%%%%%%%%%
%%%%%%%%%%%%%%%%%%%%%%%%%%%%%%%%%%%%%%%%%%%%%%%%%%%%%%%%%%%%%%%%%%%%%%%%

%%%%%%%%%%%%%%%%%%%%%%%%%%%%%%%%%%%%%%%%%%%%%%%%%%%%%%%%%%%%%%%%%%%%%%%%
%%%%%%%%%%%%%%%%%%%%%%%%%%%%%%%%%%%%%%%%%%%%%%%%%%%%%%%%%%%%%%%%%%%%%%%%
\section{Introduction}

Strongly Interacting Massive Particle (SIMPS), 
by which we will always mean neutral, stable SIMPs, are of current
interest for at least three reasons:

\begin{itemize}	

\item
They could be a dark matter constituent as suggested some time
ago by Dover, Gaisser and Steigman \fcite{dgs} and
by  Wolfram \fcite{wolfram}.
 Starkman et al.,\fcite{starkman} show SIMPs 
would be restricted to rather narrow mass ranges 
if they were to exhaust $\Omega =1$.
We will not make this
assumption and  will consider SIMPs outside the regions allowed by
the analysis of ref.~\fcite{starkman}.

\item
It is possible that the lightest SUSY particle (LSP) is strongly
interacting and hence, if R-parity is conserved, would form a 
colorless SIMP.  Possibilities, such as a $\tilde{g}g$ bound state are
discussed in ref.~\fcite{raby}.

\item
An explanation of the ultra high energy cosmic ray events
(UHECRs) proposed by Farrar, Kolb and co-workers \fcite{fketal} is 
that they are due to interactions of 
 SIMPs with a
mass below 50 GeV  and a cross section for interactions with nucleons on
the order of a few millibarns or more.

\end{itemize} 
\noindent
This summary will review two laboratory experiments that might detect
SIMPs.  More detail can be found in 
ref.~\fcite{most} and the paper on which it is based, ref~\fcite{mt}.

In Section 2 we consider the possibility of finding SIMPs bound in
ordinary nuclei by searching for anomalously heavy isotopes of high-Z
nuclei.  It is a pleasure to note that 
the accelerator mass spectrometry (AMS) group at Purdue is in the
process of performing the experiment\footnote{%
We are grateful to Professor Ephraim Fischbach for
keeping us informed as to progress on this  experiment.}
 suggested in ref.~\fcite{most}.
In Section 3, we address
the extent to which production and detection of 
SIMP--anti-SIMP ($S\overbar{S}$) pairs might
be performed at the Tevatron.

Our results, in brief, are that the AMS experiment should be sensitive to
SIMPs over a wide range of parameter space: $(\sigma_{SN},M_S)$, where
$M_S$ is  the SIMP mass and $\sigma_{SN}$ is its cross section for
scattering off nucleons.  The Tevatron, on the other hand is likely only
to produce and to detect SIMPs in a much more restricted range, but one
that includes much of the mass range for which the SIMP could be the UHECR
explanation.  It would be only fitting, since much of the work on that
possibility \fcite{fketal} was done at Fermilab, if SIMPs were to be
detected at Fermilab and we encourage those with influence in the
collaborations to explore vigorously that possibility.  Finally, we note
that we proceed without committing to a specific SIMP model.  We
parameterize the experimental predictions in terms of the two parameters
$\sigma_{SN}$ and $M_S$.

%%%%%%%%%%%%%%%%%%%%%%%%%%%%%%%%%%%%%%%%%%%%%%%%%%%%%%%%%%%%%%%%%%%%%%%%
\section{SIMPs in Nuclei}

We know a fair amount about SIMP binding in nuclei from the phenomenology
of hyper-fragments.  See, for example, Povh \fcite{povh} for a readable
review.  Based on that experience, we can write for the binding $B$ of the
SIMP in a nucleus A the relation:
\be
B=  | V_{SN}|  -\pi^2/(2\mu R^2) \quad ,
\ee
where $\mu$ is the reduced mass of the S-A system, R is the radius of the
nucleus A, and V is the S-N potential averaged over the volume of the
nucleus X.  We expect the low energy potential, $V_{SN}$, to be always 
attractive. This is true
if exchange of vacuum quantum numbers dominates.  
We assume this to be the case, and
have not found a model to the contrary.  
Under this assumption,  the SIMP can be bound
in a nucleus for which $\mu$ and $R^2\sim A^{2/3}$ are large enough to
make the kinetic energy less than the (average) magnitude of the 
attractive potential.  

From equation (1) we see  that the best chance of finding
SIMPs  is to search in high Z (large) nuclei which minimize the
kinetic energy term. 
Capture by light elements 
at the time of
cosmic nucleosynthesis has been studied in ref.~\fcite{dt}.   
Atomic Mass Spectrometer 
(AMS) searches to date
are             
reviewed in the careful study of Hemmick et al., \fcite{hemmick} where one
learns the somewhat surprising fact that previous searches have only been
conducted up to sodium ($Z=11, A=23$).  This makes the 
current Purdue AMS
experiment particularly exciting.  They are looking in gold ($Z=79,
A=197$).  

How big is the potential $V_{XN}$?  We take this as a parameter, but we
can put an approximate LOWER bound on it from the requirement that
primordial $S$ and $\overbar{S}$, left over from the early universe, not
overclose the universe so that it couldn't have continued expanding until
today (early 2000).  
The classic book of Kolb and Turner \fcite{kt} tells us
that the number density of primordial SIMPs behaves as
\be
 n_S\sim (M_S \, \sigma_{S\overbar{S}})^{-1} \quad.
\ee

Equation (2) says that too small an annihilation cross section means too
many SIMPs will be left over from the early universe, and Kolb and Turner
collect together the numerical recipes for computing how small is too
small.  We still need, however, to relate the annihilation cross section,
$\sigma_{S\overbar{S}}$ to the SIMP-nucleon cross section, $\sigma_{SN}$
and to the $S-N$ potential in Equation (1).  We make the simple ansatz
\be
 V_{SN}=V_{NN} \, (\sigma_{SN}/ \sigma_{NN})^{1/2}
\ee
\be
 \sigma_{SN}^2 =\beta \, \sigma_{NN} \, \sigma_{S\overbar{S}}
\ee
where $\beta$ should be on the order of one.  Note that $V_{SN}$ goes as
$\beta^{1/4}$ so that our results for binding will not be highly dependent
on the precision of Equation (3).

Now that we know, for each point in the $M_S$, $\sigma_{SN}$ parameter space, 
the primordial S abundance and the binding energy in nuclei, we are almost
ready to compute for our friends at Purdue, the abundance of anomalous
gold--gold with a SIMP bound in it.  First, however, we need a scenario
for how the SIMPs get bound into the gold.  Our picture is as follows:

\begin{itemize}

\item
We assume that the ratio of SIMPs to protons in the galaxy is
the same as the cosmic ratio, but that most of the SIMPs are in the
galactic halo 
({\it i.e.}, that their density distribution is $\rho \sim R^{-2}$,
where R is the distance from the galactic center), 
not in stars.  We can then calculate
the SIMP flux on the Earth, since we know that the Earth is traveling
through the galaxy with a velocity of about $200 km/s$ 
which not too different from the
galactic virial velocity.

\item
We assume that when the SIMP hits the Earth, it is slowed by
scattering with all nucleons and nuclei at a rate determined by
$\sigma_{SN}$, but can only be captured by a nucleus 
that is large enough.

\item
Gold must compete, for SIMP capture, with the most abundant
nuclei large enough to bind the SIMP.  Our comparative
estimates use, as the
most abundant elements:, 
aluminum ($A=27$), barium ($A=137$), and lead
($A=206$).

\end{itemize}
\noindent
Our procedure is then as follows:
\begin{itemize}

\item
We chose values for $M_S$ and $\sigma_{SN}$ and then determine
whether, for that point in parameter space, there is binding in gold.

\item
Assuming that there is binding, we then  
determine (a) the mean free path in Earth from the galactic
virial velocity and $\sigma_{SN}$, and (b) which of the 3 elements above is
gold's chief competitor for SIMP capture.

\item
From the ratio of the abundance of gold to its chief competitor,
the mean free path, and the average density of Earth, we then compute the
chance of a particular gold nucleus within a mean free path to capture an
incident SIMP.  Multiplying by the flux (see above) of SIMPs and the time
for which the sample being put in the AMS target has been exposed gives us
the fraction of gold nuclei in the sample that should have a SIMP if they
exist at that point in parameter space.
\end{itemize}
\noindent

Finally, we assume\footnote{%
We appreciate conversations with Professor E. T. Herrin
on searching for old exposed gold, 
and we note that SMU geologist, Dr.
Douglas Oliver, has secured such samples for the Purdue experiment.}
 that the exposure time is 10 million years
because there are regions  that are
geologically inactive over such periods and have had 
for example ``placer" gold in the
beds
of streams for a longer period than that.

The results are shown in the table.  It gives  $\log_{10}$  of
the ratio of normal to anomalous gold nuclei.  The dashes indicate 
parameter values for which there is either no binding in gold or
overclosure of the universe.  One sees that smaller values of
$\sigma_{SN}$ give larger ratios of anomalous to normal gold.  This is
because smaller values imply that only lead has a nucleus large
enough to compete with
gold for SIMP capture and because the smaller cross section means more
primordial abundance.  The important thing to take away from hours of
table study is the fact that the relative abundance entries are all
considerably higher (for anomalous to normal) than the 
limits of $10^{-20}$ that
have been set in AMS work on some of the light elements.  This provides
reason to expect that, if the SIMPs are there, the Purdue AMS people will find
them.

%%%%%%%%%%%%%%%%%%%%%%%%%%%%%%%%%%%%%%%%%%%%%%%%
%%%%%%%%%%%%%%%%%%%%%%%%%%%%%%%%%%%%%%%%%%%%%%%%
%\figtab 
%%%%%%%%%%%%%%%%%%%%%%%%%%%%%%%%%%%%%%%%%%%%%%%%
%%%%%%%%%%%%%%%%%%%%%%%%%%%%%%%%%%%%%%%%%%%%%%%%
\begin{table}[t]
\begin{center}
\begin{math}
\begin{array}{||c||c|c|c|c|c|c|c|c|c|c|c|c|c|c|c||}  \hline\hline
& 0.0005  & 0.0042   & 0.012  &  0.032 & 0.25 & 0.69 &  1.9 & 5.3 & 15 & 41 & 110 &  860  \\  \hline\hline
1.0 &     - &  - &  - &  - &  - &  - &  - & 6.3 & 8.3 & 8.7 & 12.5 &  13.4 \\ \hline
1.6 &     - &  - &  - &  - &  - &  - & 6.1 & 8.1 & 8.5 & 12.3& 12.7 &  13.6 \\ \hline
2.7 &     - &  - &  - &  - &  - & 5.9 & 7.9 & 8.3 & 12.1 & 12.5 & 12.9 &  13.8\\ \hline
4.3 &     - &  - &  - &  - & 5.7 & 7.7 & 8.1 & 11.1 & 12.3 & 12.7 & 13.1 &  14.0\\ \hline
7.1 &   - &  - &  - &  - &  7.5 & 7.9 & 10.9 & 12.1 & 12.5 & 12.9 & 13.4 &  14.2\\ \hline
12  &   - &  - &  - & 5.6 &  8.1 &8.5 & 12.2 & 12.7& 13.1 & 13.5 & 13.9 &  14.8\\ \hline
19  &   - &  - &  - &7.5 &  8.3 &11.3 & 12.5 & 12.9 &13.3 & 13.8 & 14.2 & 15.0\\ \hline
31  &   - &  - & 7.4 &7.8 & 8.6 & 12.4&12.8 & 13.2 & 13.6& 14.1 & 14.5 &  15.3\\ \hline
50  &   - & 5.7 &7.7 &8.1&11.5& 12.7 & 13.1 & 13.6 & 14.0 &14.4 &14.8 &15.7\\ \hline
81  & 5.7 &7.7 &8.1 &8.5 &11.9 &13.1 &13.5 &14.0 &14.4 &14.8 &15.2 &16.1\\ \hline
132 & 7.7 &8.1 &8.5 &8.9 &12.2 &13.5 &13.9 &14.3 &14.7 &15.2 &15.6 &16.4\\ \hline
220 & 8.0 &8.4 &8.9 &9.3 &12.6 &13.9 &14.3 &14.7 &15.1 &15.5 &16.0 &16.8\\ \hline
350 & 8.4 &8.8 &9.3 &9.7 &13.8 &14.3 &14.7 &15.1 &15.5 &15.9 &16.4 &17.2\\ \hline
570 & 8.8 &9.2 &9.7 &10.1 &14.3 &14.7 &15.1 &15.5 &15.9 &16.4 &16.8  &17.6\\ \hline
930 & 9.3 &9.7 &10.1 &10.5 &14.7 &15.1 &15.5 &16.0 &16.4 &16.8 &17.2 &18.1\\  \hline\hline
\end{array}
\end{math}
\vskip 10pt
\caption{
$M_X$ (vertical) is in units of GeV, and $\sigma_{XN}$ (horizontal) is in units of $mb$.  
 Table entries are $\log_{10}(1/f)$, and
 the $-$ indicates those cases for which $X$ does not bind at all.
}
\end{center}
\end{table}
%%%%%%%%%%%%%%%%%%%%%%%%%%%%%%%%%%%%%%%%%%%%%%%%
%%%%%%%%%%%%%%%%%%%%%%%%%%%%%%%%%%%%%%%%%%%%%%%%

%%%%%%%%%%%%%%%%%%%%%%%%%%%%%%%%%%%%%%%%%%%%%%%%
%%%%%%%%%%%%%%%%%%%%%%%%%%%%%%%%%%%%%%%%%%%%%%%%

%%%%%%%%%%%%%%%%%%%%%%%%%%%%%%%%%%%%%%%%%%%%%%%%%%%%%%%%%%%%%%%%%%%%%%%%
\section{SIMPs at Fermilab}

\figthree

Next we consider $S\overbar{S}$ production at the Tevatron.  Since we are
talking neutral SIMPs, we expect little or no signal in the central
tracker and in the electromagnetic calorimeter.  However, in the hadron
calorimeter, we expect to 
detect SIMP signals  if $\sigma_{SN}$ is large enough.
The detection of SIMPs is possible if one triggers on two 
relatively back-to-back
hadron calorimeter showers, accompanied by little else.  We will use 10
GeV for the minimum size showers for which such triggering might be done. 
Our task now is to determine:

\begin{itemize}
\item 
For what values of \{$M_S,\sigma_S$\} will the SIMP interact in the steel
plates of the hadron calorimeter?

\item 
For what values of these parameters will we get 
calorimeter showers greater than 
10 GeV or more?

\item 
Can one recognize a SIMP shower if one sees one?

\item 
How many such events should we expect?
\end{itemize}
\noindent
 First we look at the region of parameter space for which there will be
interaction.  The minimum annihilation cross section permitted by the
cosmology argument is \hbox{$\sim 3\times 10^{-13}barns$,} which corresponds
through Equation (3) to about a microbarn for the S-N cross section.  
SIMPs with such small cross sections won't shower in 1
meter of steel, but for a higher cross section of a few millibarns, we would
expect 10 or more interactions with the $10^{27}nucleons/cm^2$  in the 1
meter. 

To estimate the energy we expect in a shower resulting
from a SIMP interaction in the steel plates of a hadron calorimeter 
we use a cosmic ray rule of thumb kindly provided by G. Yodh\footnote{%
This useful approximation from Professor Gaurang Yodh made
the whole trip to Paris (where the conversation took place) well
worthwhile (and the food was OK too).}
who says that, in a high energy strong interaction, about half the
center of mass energy goes into inelasticity.  In the figure, we give the
(laboratory) energy released in the calorimeter as we vary the mass and
energy of the SIMP; the straight lines are constant shower energies.  One
sees that the bigger the SIMP lab energy, the greater a SIMP mass will
result in a given shower energy.

Consider now the question of whether we would recognize a SIMP shower
if we saw one.  The background for SIMP showers would likely be neutron
showers and $K$ decays.  
The distinguishing feature would be shower opening angle.  A
pion moving transverse in the c.m. system would have a lab angle given
by $\tan\theta = 1/\gamma$.
Comparing the angle for a SIMP with that from a neutron of the same energy, 
the SIMP shower should be wider by roughly the ratio of the masses. 

Finally, we turn to the number of SIMP pairs the Tevatron might produce.
We scale the (known) production rate of jets by the ratio of the S-N
cross section to that of Meson-N, which we take to be on the order of 30
millibarns.  So long as the SIMP energy is a few times its mass, we don't
worry about phase space suppression.  We assume conservatively a cross
section of about $3 pb$ for any one parton in the region $E>200GeV$.  This
implies about  6000 events at the Tevatron Run II.  The
estimate of \fcite{fketal} is that the Nucleon-UHECR cross section needs
to be over a tenth the Meson-Nucleon cross section, so we estimate 
over 600 events in the Tevatron Run II if SIMPs are the explanation for the UHECR
events.

%%%%%%%%%%%%%%%%%%%%%%%%%%%%%%%%%%%%%%%%%%%%%%%%%%%%%%%%%%%%%%%%%%%%%%%%
\section{Summary}   

From the Table we see that there is SIMP binding in gold for
$M_S^2\sigma_{SN}>5mb\,GeV^2$, and that AMS experiments sensitive to one
part in $10^{20}$ can detect the existence of SIMPs of mass less
than a TeV, while the region of interest for explaining UHECRs can be
explored with a sensitivity of one part in $10^{16}$ or less.  Looking for
SIMPs at the Tevatron is more difficult, but over half the region of
interest for explaining UHECRs could be searched in the upcoming Run~II
by looking for (wide) back to back jets with no signal in the
central tracker or EM calorimeter.

%%%%%%% Acknowledgments: 
We thank 
D.~Berley, 
% K.~Brockett,
K.~De,
D.~Dicus,
M.A.~Doncheski,
R.~Ellsworth, 
G.~Farrar, 
T.~Ferbel, 
E.~Fischbach, 
E.T.~Herrin,
G.~Landsberg,
D.~Oliver, 
D.~Rosenbaum,
R.~Scalise,
and
G.~Yodh.
 The work of RNM has been supported by
the National Science Foundation grant under no. PHY-9802551 .  
The work of Olness, Stroynowski, and Teplitz is supported by DOE.

%%%%%%%%%%%%%%%%%%%%%%%%%%%%%%%%%%%%%%%%%%%%%%%%%%%%%%%%%%%%%%%%%%%%%%%%

%%%%%%%%%%%%%%%%%%%%%%%%%%%%%%%%%%%%%%%%%%%%%%%%%%%%%%%%%%%%%%%%%%%%%%%%%%%%%%%%

%INDEX%%%%%%%%%%%%%%%%%%%%%%%%%%%%%%%%%%%%%%%%%%%%%%%%%%%%%%%%%%%%%%%
% Please check with the editor of your book whether he plans to
% include a "mutual" subject index - if so, please code your entries
% in the standard syntax. For your own purposes you may print your
% "personal" index by using the following commands:
%
%\clearpage
%\addcontentsline{toc}{section}{Index}
%\flushbottom
%\printindex
%%%%%%%%%%%%%%%%%%%%%%%%%%%%%%%%%%%%%%%%%%%%%%%%%%%%%%%%%%%%%%%%%%%%%

\end{document}